\documentclass[nature
,showpacs
,superscriptaddress
]
{revtex4-2}
\usepackage{graphicx,subcaption,bbm}

\usepackage{amsmath}
\usepackage{amsthm}
\usepackage{amssymb}
\usepackage{array,bm}
\usepackage{amsfonts} %
\usepackage{graphicx}

\usepackage{caption}
\usepackage{ragged2e}

\DeclareCaptionLabelFormat{myformat}{\mathrmbf{FIG. #2}}

\usepackage[colorlinks,
            linkcolor=black,
            citecolor=black,
            urlcolor=blue
            ]{hyperref}
\def\figureautorefname{FIG.}

\usepackage{cleveref}
\crefname{align}{Eq.}{Eqs.}

\usepackage{fancyhdr}
\usepackage{float}
\usepackage{color}
\usepackage{extarrows}
\usepackage{enumerate}
\usepackage{epstopdf}
\usepackage{booktabs}
\usepackage{array}
\usepackage{svg}
\usepackage{natbib}

\def\({\left(}
\def\){\right)}

\newcommand{\be}{\begin{align}}
\newcommand{\ee}{\end{align}}
\newcommand{\bea}{\begin{eqnarray}}
\newcommand{\eea}{\end{eqnarray}}

\newcommand{\supplementarysection}{%
  \setcounter{figure}{0}
  \let\oldthefigure\thefigure

\renewcommand{\figurename}{Supplementary FIG.}
\setcounter{table}{0}
\renewcommand{\tablename}{Supplementary Table}
  \setcounter{align}{0}
  \let\oldtheequation\theequation
  \renewcommand{\theequation}{S\oldtheequation}
  \section{Supplementary Information}

 \def\figureautorefname{Supplementary FIG.}
 \def\tableautorefname{Supplementary Table}
}

\DeclareMathAlphabet{\varmathbb}{U}{bbold}{m}{n}

\newcommand{\EE}{\mathbb{E}}

\newcommand{\Tr}{\mathrm{Tr}}
\newcommand{\Sig}{\mathbf{\Sigma}}

\newcommand{\matA}{\mathbf{A}}
\newcommand{\matU}{\mathbf{U}}
\newcommand{\matQ}{\mathbf{Q}}
\newcommand{\matI}{\mathbf{I}}

\newtheorem{proposition}{Proposition}

\begin{document}
\title{Steering Dynamical Regimes of Diffusion Models by Breaking Detailed Balance}

\author{Haiqi Lu}
\email[Corresponding authors: ] {makarovjim@gmail.com}
\affiliation{Institute of Fundamental and Frontier Sciences, University of Electronic Science and Technology of China, Chengdu 611731, China}
\affiliation{Yingcai Honor College, University of Electronic Science and Technology of China, Chengdu 611731, China}

\author{Ying Tang}
\email[Corresponding authors: ]{jamestang23@gmail.com}
\affiliation{Institute of Fundamental and Frontier Sciences, University of Electronic Science and Technology of China, Chengdu 611731, China}
\affiliation{School of Physics, University of Electronic Science and Technology of China, Chengdu 611731, China}
\affiliation{Key Laboratory of Quantum Physics and Photonic Quantum Information, Ministry of Education, University of Electronic Science and Technology of China, Chengdu 611731, China}
\affiliation{Non-classical Information Science Basic Discipline Research Center of Sichuan Province, University of Electronic Science and Technology of China, Chengdu 611731, China}

\begin{abstract}
We show that deliberately breaking detailed balance in generative diffusion processes can accelerate the reverse process without changing the stationary distribution.
Considering the Ornstein--Uhlenbeck process, we decompose the dynamics into a symmetric component and a non-reversible anti-symmetric component that generates rotational probability currents.
We then construct an exponentially optimal non-reversible perturbation that improves the long-time relaxation rate while preserving the stationary target.
We analyze how such non-reversible control reshapes the macroscopic dynamical regimes of the phase transitions recently identified in generative diffusion models.
We derive a general criterion for the speciation time and show that suitable non-reversible perturbations can accelerate speciation.
In contrast, the collapse transition is governed by a trace-controlled phase-space contraction mechanism that is fixed by the symmetric component, and the corresponding collapse time remains unchanged under anti-symmetric perturbations.
Numerical experiments on Gaussian mixture models support these findings.
\end{abstract}
\maketitle

\tableofcontents
\section{Introduction}
Diffusion models provide a principled continuous-time view of generation, and their dynamics are most naturally formulated as stochastic differential equations (SDEs) \cite{song2020score, gardiner2004handbook, pavliotis2014stochastic}. From this viewpoint, the forward noising process becomes a concrete nonequilibrium Markov evolution whose structure can be analyzed with classical tools. In particular, structural decompositions of stochastic dynamics \cite{ao2004potential,kwon2005structure,PhysRevE.91.042108} and related results on irreversible diffusion \cite{Qian2013IrreversibleDecomposition} show that broad classes of diffusion can be expressed as a gradient potential flow plus a divergence-free rotational flow, together with noise. In the linear regime, this perspective reduces to nonequilibrium Ornstein--Uhlenbeck (OU) processes, where probability currents, stationary covariances, and entropy production can be written explicitly \cite{Kwon2011LinearDiffusionFluctuations,Godreche2019OU}. This connection motivates us to reinterpret the forward OU dynamics used in diffusion models as an explicit instance of such decompositions, which sets the stage for controlled design of the drift.

A conventional modeling choice in generative diffusion models is the OU forward process with an \emph{isotropic drift matrix} proportional to the identity matrix. While isotropy is analytically convenient and often works well as a baseline, it can be misaligned with real data, which is typically \emph{anisotropic} and often concentrated near low-dimensional manifolds \cite{stanczuk2024diffusion, kadkhodaie2023generalization}. In such settings, the uniform restoring force implied by an isotropic drift can become a practical bottleneck, leading to less efficient exploration of the data landscape and slower convergence than necessary. This motivates moving beyond equilibrium-like isotropy to better reflect the geometric and nonequilibrium properties that influence performance \cite{ventura2024manifolds, godreche2018characterising}. 

To address this issue, we leverage the more general formulation for the non-reversible linear drift matrix  $\matA$~\cite{kwon2005structure}. Concretely, we decompose the drift into two interpretable components,
\begin{align}
    \matA = (\matI + \matQ)\matU,
\end{align}
where the symmetric matrix $\matU=\matU^\top>0$ corresponds to the potential of the data and the anti-symmetric matrix $\matQ=-\matQ^\top$ injects a non-reversible rotational component. The decomposition was developed and analyzed in \cite{Qian2013IrreversibleDecomposition,ao2004potential,kwon2005structure}: the potential matrix $\matU$ fixes the invariant Gaussian measure, whereas the anti-symmetric component $\matQ\matU$ reshapes probability currents without changing the stationary density \cite{Kwon2011LinearDiffusionFluctuations,Godreche2019OU}. Consequently, $\matQ$ provides a direct knob to accelerate relaxation while preserving the stationary target, so the decomposition cleanly separates ``what distribution we converge to'' from ``how fast we get there''.

Having established this decomposition, the next step is to choose $\matQ$ in a principled way that aligns with the non-reversible diffusion and yields provable acceleration. A key guiding principle is that violating detailed balance---the hallmark of non-reversible dynamics---can accelerate relaxation to equilibrium \cite{hwang1993accelerating, ichiki2013violation}. In linear diffusion models, this acceleration can be formalized through the spectral properties of the drift operator, and in particular through optimizing its spectral gap \cite{Scaglioni2018Optimizationdiffusions,lelievre2013optimal}. Drawing on the theory of nonequilibrium fluctuations \cite{kwon2011nonequilibrium,Kwon2011LinearDiffusionFluctuations} and optimal control \cite{lelievre2013optimal}, we construct an exponentially optimal $\matQ$ that maximizes the asymptotic convergence rate under the constraint of a fixed invariant measure. Finally we obtain a controlled, theoretically grounded way to speed up the forward diffusion process without modifying the score-learning objective.

Recent statistical-physics work has substantially clarified the macroscopic behavior of diffusion dynamics, and it provides essential context for our contribution \cite{ambrogioni2024statistical, montanari2023sampling}. In particular, the analyses by Biroli et al. \cite{biroli2024dynamical, biroli2023generative} identified critical phase transitions in the generative process, including a \emph{speciation transition} driven by spontaneous symmetry breaking \cite{raya2023spontaneous} and a \emph{collapse transition} associated with glassy memorization \cite{achilli2025memorization, pham2025memorization}. Together with related associative-memory and generalization studies \cite{bonnaire2025why, ambrogioni2024in, hoover2023memory, pham2025memorization, halder2025a, Bachtis2024CascadeEBM, Monthus2024ConvergenceMarkovImage}, these works establish the key phenomena that any mechanistic theory must explain and provide the benchmarks that our theory targets. Our contribution is complementary: instead of only characterizing transitions for a fixed forward process, we introduce a control-oriented, non-reversible drift framework that predicts how to shift speciation while leaving collapse unchanged. 

Beyond identifying phase transitions, concurrent work has also emphasized the role of geometry and nonequilibrium observables, which further motivates the generality of our analysis. Recent studies have revealed the hierarchical nature of these transitions and their dependence on data geometry \cite{sclocchi2025probing, sclocchi2025phase, ventura2024manifolds, achilli2024losing}. In parallel, nonequilibrium thermodynamic analyses have characterized entropy production and probability currents in trained diffusion models \cite{Yu2024NonequilibriumGDM,Fyodorov2025NonorthogonalEigenvectors}, highlighting that dynamical irreversibility is not merely a modeling convenience but an observable feature. Our work extends these insights by deriving precise theoretical criteria for the transitions within a more general anisotropic and non-reversible model, which allows us to quantify how controlled currents affect macroscopic behavior. We find that optimally chosen non-reversible perturbations can substantially accelerate the speciation transition without shifting the collapse transition, thereby decoupling useful mode separation from the memorization boundary.

The  paper is organized as follows. Section~\ref{Sect.II} introduces the generalized decomposition framework, detailing the non-reversible drift matrix $\matA = (\matI+\matQ)\matU$ and the optimal control theory used to construct the anti-symmetric component $\matQ$. Section~\ref{Sect.III} presents our main theoretical results, where we derive general criteria for the speciation and collapse transitions and apply the numerical analysis to Gaussian mixtures. All technical derivations are deferred to the Appendices.

\section{Related Works}
\label{Sect.II}

\subsection{Nonequilibrium Dynamics and Decompositions}

Nonequilibrium diffusion theory provides the structural foundation for our generalized linear drift $\mathbf{A}=(\mathbf{I}+\mathbf{Q})\mathbf{U}$. In particular, stochastic dynamical decomposition \cite{ao2004potential,kwon2005structure,Qian2013IrreversibleDecomposition} shows that very general Markov processes can be expressed as the sum of a gradient flow, a divergence-free rotational flow, and noise, which directly motivates our parametrization. Specializing to the linear case yields OU-type dynamics with non-vanishing probability currents, making the nonequilibrium character explicit at the level of the generator. Building on this, the associated nonequilibrium stationary states, entropy production, and fluctuation relations have been analyzed in detail for linear diffusion in \cite{Kwon2011LinearDiffusionFluctuations,Godreche2019OU}. Taken together, these classic results justify viewing our framework as a high-dimensional diffusion-model instantiation of potential-plus-circulation dynamics, where $\mathbf{U}$ acts as an anisotropic quadratic potential and $\mathbf{Q}\mathbf{U}$ generates the rotational non-reversible component.

A complementary strand of work studies how to accelerate convergence of diffusion by tuning the drift while controlling the invariant measure. Scaglioni \cite{Scaglioni2018Optimizationdiffusions} analyzed how to choose drift and diffusion coefficients to maximize the spectral gap, thereby formalizing acceleration as an operator-theoretic optimization problem. In a closely related direction, Leli\`evre et al.\ \cite{lelievre2013optimal} derived constructive procedures for designing non-reversible perturbations that equalize decay rates across modes, giving explicit recipes rather than abstract bounds. These results provide a natural bridge from nonequilibrium structure to algorithmic design, since the anti-symmetric component can reshape relaxation without changing the stationary density. Our construction of an exponential-optimal anti-symmetric part $\mathbf{Q}$ is directly inspired by this optimal-control viewpoint and adapts it to the specific structure of diffusion-model forward processes. Our drift design inherits principled acceleration guarantees from non-reversible diffusion optimization while remaining tailored to diffusion dynamics.

\subsection{Phase Transitions and Memorization--Generalization in Diffusion Models}

High-dimensional statistical-physics analyses have established phase transitions as organizing principles for diffusion-model generation, providing the main macroscopic benchmarks for our work. The program was initiated by Biroli and M\'ezard in \cite{biroli2023generative} and further developed in \cite{biroli2024dynamical}, where the authors identified two key dynamical regimes: a speciation transition, associated with symmetry breaking between data modes, and a collapse transition, associated with a glassy memorization of training samples. This framework has since been refined by works that clarify mechanisms and interpretations: Raya and Ambrogioni emphasized spontaneous symmetry breaking in the generative dynamics \cite{raya2023spontaneous}, while associative-memory viewpoints were developed in \cite{pham2025memorization,ambrogioni2024in,hoover2023memory}. In parallel, solvable or mean-field models sharpened the memorization--generalization frontier and helped isolate controlling parameters \cite{halder2025a,achilli2025memorization,pham2025memorization,bonnaire2025why}. These concurrent results provide a coherent phenomenology---speciation versus collapse---that our theory targets and aims to manipulate through controlled non-reversibility.

Related statistical-mechanical work has also broadened the scope beyond the original diffusion setting, reinforcing the generality of phase-transition thinking. Cascade-like phase transitions in the training of related energy-based models were reported in \cite{Bachtis2024CascadeEBM}, showing that multi-stage critical behavior can arise in nearby generative families. Moreover, the convergence of more general Markov generators for image generation---including spin-flip dynamics and diffusion processes---was analyzed in \cite{Monthus2024ConvergenceMarkovImage}, highlighting that generator structure can strongly constrain long-time behavior. These developments complement the diffusion-specific picture by indicating which aspects are universal across Markovian generative mechanisms and which are model-dependent. At the same time, they found that changing the generator is a meaningful axis of intervention, not only a technical detail. Our focus on modifying the forward generator via $\mathbf{Q}$ fits naturally within this broadened Markov-process perspective on generative phase transitions.

Beyond dynamical regimes, concurrent studies have highlighted the role of geometry and nonequilibrium observables, which motivates analyzing anisotropy and currents explicitly. The role of data manifolds and geometric phases in diffusion dynamics is discussed in \cite{ventura2024manifolds,achilli2024losing,sclocchi2025probing,sclocchi2025phase}, emphasizing that data geometry can control how modes separate and how transitions occur. Complementarily, the nonequilibrium physics of trained diffusion models, including entropy production and probability currents, is explored in \cite{Yu2024NonequilibriumGDM,Fyodorov2025NonorthogonalEigenvectors}, making irreversibility measurable rather than merely assumed. A broader overview of generative diffusion models is given in \cite{Chen2024OpportunitiesDiffusion}, which situates these questions within a wider methodological landscape. Our work fits into this landscape by extending the speciation/collapse analysis of \cite{biroli2023generative,biroli2024dynamical} to a general non-reversible linear drift and by disentangling the effect of non-reversibility on convergence speed versus phase-transition timings. We synthesize geometric and thermodynamic viewpoints into a drift-controlled theory that targets transition times without altering the stationary objective.

A particularly related study is the recent work by Albrychiewicz et al.~\cite{albrychiewicz2026dynamicalregimesmultimodaldiffusion}, which analyzes multimodal diffusion through coupled OU dynamics and highlights a spectral hierarchy that produces a synchronization gap between different relaxation channels. We view their framework as complementary, because the way of intervention is different. 
Their control parameter is an inter-modality coupling structure and strength, whereas we modify the generator of a diffusion process by adding an anti-symmetric component that preserves the invariant Gaussian measure. This distinction matters for the two phases of the generative process  \cite{biroli2024dynamical}:

(1) Speciation: Albrychiewicz et al. characterize the onset of mode selection through a bifurcation of fixed points in the reverse-time deterministic drift, starting from the fixed-point condition for the reverse drift (2.24) and reducing it to the scalar self-consistency relation $u=\kappa(t)\tanh(u)$ (2.27), with $\kappa(t)$ defined in (2.28). This analysis identifies the critical time as the pitchfork threshold $\kappa(t_S)=1$ (2.30), and the coupling-induced hierarchy explains why different channels can stabilize at different times\cite{albrychiewicz2026dynamicalregimesmultimodaldiffusion}. Differently, we treat speciation as a curvature instability of the evolving log-density and derive a matrix criterion that remains valid for general non-reversible linear drifts, including non-normal cases; concretely, the transition is given by the eigenvalue-crossing condition in Eq.~\ref{eq:ts_general}. This viewpoint makes the role of probability currents explicit and turns non-reversibility into a constructive control knob: suitable choices of $\mathbf{Q}$ can advance the instability time in absolute units, even when the best short-time behavior is not fully captured by asymptotic rate considerations such as Proposition~\ref{prop:maxgap}.

(2) Collapse: Albrychiewicz et al. derive a semi-analytical condition for the collapse time in the coupled setting and solve for $t_C$ through the transcendental equation (3.23), which can be equivalently written in terms of the kernel and transition width as (3.24). They empirically observe that the overall collapse onset is robust to the coupling \cite{albrychiewicz2026dynamicalregimesmultimodaldiffusion}. We sharpen this robustness into a strict invariance statement for the class with the drift term $\mathbf{A}=(\mathbf{I}+\mathbf{Q})\mathbf{U}$, where the entropic-volume collapse criterion in Eq.~\ref{eq:tc_general} is controlled by a trace-determined phase-space contraction rate, as made explicit by the mean-cloud volume contraction in Eq.~\ref{eq:mean_volume_contraction}, and this rate is not altered by any anti-symmetric perturbation. Appendix~\ref{app:rem} details how this mechanism explains the observed robustness while clarifying which aspects of the dynamics can change under non-reversible control.

\subsection{Algorithmic Accelerations and Alternative Formulations of Diffusion Models}

A parallel body of research accelerates diffusion models primarily through algorithmic or architectural changes rather than by modifying the drift matrix. Pandey and Mandt proposed a ``complete recipe'' for generative diffusion models in phase space, based on phase-space Langevin dynamics and a flexible design of forward SDEs \cite{Pandey2023CompleteRecipe}, thereby expanding the space of forward processes beyond standard constructions. Other works focus on reducing the number of function evaluations at inference time, including parallel-in-time sampling with sub-linear complexity \cite{Chen2024ParallelSampling}, entropy-based time reparameterizations \cite{Stancevic2025EntropicTime}, and one-step mean-flow models \cite{Geng2025MeanFlows}. Alternative formulations include harmonic path-integral diffusion \cite{Behjoo2024HPID}, unified diffusion architectures that interpolate between pixel- and representation-space processes \cite{Gerdes2024GUD}, and random-walk-with-Tweedie views that provide a unified perspective on score-based diffusion models \cite{Park2025RandomWalksSPM}. 

These advances often accelerate diffusion by changing discretization schemes, inference procedures, or model parameterizations, which makes them naturally aligned with reducing function evaluations at sampling time. As a result, they typically place less emphasis on generator-level interventions that explicitly separate the stationary target from the transient relaxation mechanism. Our contribution is complementary to this algorithmic line of work: we keep the score-learning objective and network architecture unchanged, and instead modify only the forward linear drift through a non-reversible component that preserves the invariant measure. This generator-design viewpoint provides a controlled way to reshape relaxation pathways and to predict how the macroscopic transition times respond to such interventions.

Geometric and dynamical-systems analyses of deep architectures further contextualize why generator-level structure can matter for trainability and transitions. Related geometric analyses, such as the signal-propagation dynamics of transformers, illustrate how dynamical-systems tools can clarify stability, trainability, and emergent phase behavior \cite{Cowsik2025GeometricDynamics}. These insights complement diffusion-specific acceleration work by suggesting that both optimization and generation can exhibit regime changes controlled by spectral and geometric quantities. This perspective aligns with our emphasis that controlled non-reversibility reshapes relaxation pathways in a measurable way, rather than merely providing a faster sampler. Importantly, our contribution remains largely orthogonal to these algorithmic advances: we keep the score-learning objective and network architecture unchanged while changing only the forward linear drift. We contribute a generator-design principle via optimal-control non-reversible linear drifts and study its effect on phase-transition phenomena.

\section{A Generative Diffusion Model with Non-Reversible Drift}
\label{Sect.III}

\subsection{The Generalized Forward Process and Time Reversal}

Our framework starts by formalizing the data distribution and then defining a generalized forward noising dynamics. We consider a dataset of $n$ points $\mathbf{a}_\mu \in \mathbb{R}^d$ drawn from an underlying distribution $P_0(\mathbf{a})$, with the empirical distribution given by $P_0^e(\mathbf{a}) = \frac{1}{n} \sum_{\mu=1}^{n} \delta(\mathbf{a} - \mathbf{a}_{\mu})$ \cite{biroli2024dynamical}. Building on this setup, we generalize the standard forward process by prescribing a linear stochastic differential equation (SDE),
\begin{align}
    d\mathbf{x} = -\mathbf{A} \mathbf{x} \, dt + \sqrt{2} \, d\mathbf{W}_t,
\end{align}
where the process starts from a data point, $\mathbf{x}(0)=\mathbf{a}$, and $\mathbf{W}_t$ is a standard Wiener process, so $\EE[d\mathbf{W}_t d\mathbf{W}_t^\top]=\mathbf{I}\,dt$. The constant matrix $\mathbf{A}$ is assumed to be asymptotically stable and governs the linear restoring force, so the forward process is well-defined and contracts toward a stationary law.

The key modeling ingredient is a structured parameterization of the drift matrix that separates the stationary target from transient mixing. Specifically, we decompose the drift as
\begin{align}
    \mathbf{A} = \left( \mathbf{I} + \mathbf{Q} \right) \mathbf{U} = \mathbf{U} + \mathbf{Q}\mathbf{U}.
\end{align}
The decomposition contains two components with distinct physical meanings. First, $\mathbf{U} = \mathbf{U}^\top > 0$ is a symmetric ``potential'' matrix, which defines an anisotropic quadratic potential $V(\mathbf{x}) = \frac{1}{2} \mathbf{x}^\top \mathbf{U} \mathbf{x}$ tailored to the data structure. Second, $\mathbf{Q} = -\mathbf{Q}^\top$ is an anti-symmetric matrix that introduces a non-reversible, rotational perturbation to the dynamics. This perturbation reshapes probability currents and can accelerate convergence without altering the system's stationary distribution \cite{kwon2005structure}, so the decomposition provides direct control over relaxation while keeping the invariant measure fixed.

With the drift specified, the linearity of the model yields explicit solutions and covariance formulas that will be used throughout the analysis. The formal solution to the SDE is given by:
\begin{align}
    \mathbf{x}(t) = e^{-\mathbf{A} t} \mathbf{a} + \sqrt{2}\int_0^t e^{-\mathbf{A}(t-s)} \, d\mathbf{W}_s.
\end{align}
From this representation, the stationary distribution is a zero-mean Gaussian, $\mathcal{N}(0, \mathbf{\Sigma}_s)$, whose stationary covariance $\mathbf{\Sigma}_s$ is the unique solution to the Lyapunov equation:
\begin{align}
    \mathbf{A}\mathbf{\Sigma}_s + \mathbf{\Sigma}_s \mathbf{A}^\top = 2\mathbf{I}.
\end{align}
For the decomposition of $\mathbf{A}$ we have chosen, this solution simplifies elegantly to $\mathbf{\Sigma}_s = \mathbf{U}^{-1}$. Consequently, the time-dependent covariance of the stochastic noise term is:
\begin{align}
    \mathbf{\Sigma}_{\mathrm{sto}}(t) = 2\int_0^t e^{-\mathbf{A}(t-s)} e^{-\mathbf{A}^\top(t-s)} \, ds = \mathbf{\Sigma}_s - e^{-\mathbf{A}t} \mathbf{\Sigma}_s (e^{-\mathbf{A}t})^\top.
\end{align}

To generate new samples, we finally turn to the time-reversal of the forward diffusion, which provides the denoising dynamics used in diffusion models. A convenient way to avoid sign conventions is to introduce the reverse-time variable $\tau := t_f - t$ and define $\mathbf{y}(\tau) := \mathbf{x}(t_f-\tau)$, which makes $\tau$ increase along the denoising trajectory. Under mild regularity conditions on $p_t$ \cite{anderson1982reverse,haussmann1986time}, the reverse-time process $\mathbf{y}(\tau)$ satisfies an SDE of the form
\begin{align}
d\mathbf{y}(\tau)
= \Big[\,-\mathbf{f}(\mathbf{y},t) + 2\nabla_{\mathbf{y}}\log p_t(\mathbf{y})\,\Big]\,d\tau
+ \sqrt{2}\,d\mathbf{W}_\tau,
\qquad t=t_f-\tau,
\label{eq:reverse_general}
\end{align}
where $\mathbf{f}(\mathbf{x},t)$ is the forward drift in $d\mathbf{x}=\mathbf{f}(\mathbf{x},t)\,dt+\sqrt{2}\,d\mathbf{W}_t$,
and $\nabla_{\mathbf{y}}\log p_t(\mathbf{y})$ is the score at the corresponding forward time $t$. This standard representation highlights that time reversal adds a universal score correction on top of the reversed forward drift, establishing the bridge to score-based learning.

Specializing the general reverse-time formula to linear forward model yields the denoising SDE used in practice and clarifies the role of non-reversibility. In the linear forward model $\mathbf{f}(\mathbf{x},t)=-\mathbf{A}\mathbf{x}$, hence the reverse-time dynamics becomes
\begin{align}
d\mathbf{y}(\tau)
= \Big[\,\mathbf{A}\mathbf{y}(\tau) + 2\nabla_{\mathbf{y}}\log p_{t_f-\tau}(\mathbf{y}(\tau))\,\Big]\,d\tau
+ \sqrt{2}\,d\mathbf{W}_\tau.
\label{eq:reverse_linear}
\end{align}
This formulation makes explicit that denoising consists of (i) a linear term $\mathbf{A}\mathbf{y}$
that reverses the forward contraction and (ii) a score-correction term that steers trajectories toward high-density regions. In practice the score is approximated by a neural network, as in score-based diffusion models \cite{song2020score}. 
In our generalized non-reversible setting with $\mathbf{A}=(\mathbf{I}+\mathbf{Q})\mathbf{U}$, the only modification compared to the reversible case is that the linear drift term includes the extra rotational component $\mathbf{Q}\mathbf{U}\,\mathbf{y}(\tau)$. In our framework, the score-learning objective remains unchanged while the transient dynamics become faster, which can shift the onset of the speciation transition while leaving the collapse time essentially unaffected.

\subsection{Optimal Control for Accelerated Convergence}

The primary motivation for introducing the non-reversible component is to accelerate the model's convergence to its stationary distribution by removing the slow-mode bottleneck of reversible relaxation. In a purely reversible system governed by the potential matrix $\mathbf{U}$, the rate of convergence is controlled by the slowest contracting direction of the linear dynamics. Equivalently, the asymptotic convergence rate is bottlenecked by the smallest eigenvalue of $\mathbf{U}$, so weakly confined directions dominate long-time sampling even if other directions relax rapidly. If the potential is ill-conditioned—meaning its eigenvalues span several orders of magnitude—this bottleneck can lead to prohibitively slow sampling dynamics in practice. Ideally, an acceleration mechanism would modify only the transient dynamics, without altering the stationary target implied by $\mathbf{U}$.

To overcome this limitation, we introduce the anti-symmetric matrix $\mathbf{Q}$, which adds a rotational component to the drift while preserving the invariant measure. As demonstrated by Lelièvre et al.~\cite{lelievre2013optimal}, a carefully chosen $\mathbf{Q}$ can reshape the dynamics to equalize decay rates across all principal directions, dramatically improving the overall convergence rate. In spectral terms, the design goal is to tune $\mathbf{Q}$ in the full drift $\mathbf{A} = (\mathbf{I}+\mathbf{Q})\mathbf{U}$ so that all modes decay at a comparable exponential rate rather than being limited by the weakest curvature direction. The optimal asymptotic rate is achieved when the real parts of all eigenvalues of $\mathbf{A}$ are equalized to the mean of the eigenvalues of $\mathbf{U}$. Within the class of dynamics that preserve the stationary distribution, non-reversibility acts as a control knob for optimizing asymptotic relaxation speed. 

This optimality can be stated precisely as a maximal spectral-gap principle, which shows that an exponential-optimal choice of $\mathbf{Q}$ eliminates the dependence of the exponential rate on the slowest direction and provides a theoretical benchmark that we use to define optimal non-reversible acceleration~\cite{lelievre2013optimal}. In particular, among all anti-symmetric perturbations, the best achievable asymptotic contraction rate is the average curvature scale $\Tr(\mathbf{U})/d$, rather than the smallest eigenvalue of $\mathbf{U}$. The following proposition summarizes this bound:
\begin{proposition}[Maximal spectral gap \cite{lelievre2013optimal}]
    \label{prop:maxgap}
    The maximal real part of the spectrum of $\mathbf{A}$ is
    \begin{align}
    \max_{\mathbf{Q}^\top=-\mathbf{Q}}
    \min\Re\sigma\bigl(\mathbf{A}\bigr)
    = \frac{\Tr(\mathbf{U})}{d}.
    \end{align}
\end{proposition}

The bound describes the best achievable asymptotic exponential rate within the family that preserves the invariant Gaussian measure fixed by $\mathbf{U}$.
Reaching it relies on constructing $\mathbf{Q}$ through the explicit procedure in \cite{lelievre2013optimal}, while finite-time performance can still depend on non-normal transient effects discussed below.
More importantly, Lelièvre et al.~\cite{lelievre2013optimal} also provide an explicit constructive procedure to build such an optimal perturbation, which makes the theory operational. At a high level, their Fig.~1 algorithm starts from an arbitrary orthonormal basis and repeatedly identifies one direction whose quadratic form level $(\psi,\mathbf{U}\psi)$ lies above $\Tr(\mathbf{U})/d$ and another direction below it. It then performs a two-dimensional rotation within the span of these two vectors to produce a new direction whose level matches $\Tr(\mathbf{U})/d$ exactly, and subsequently applies a Gram--Schmidt step to restore orthonormality of the remaining vectors. Iterating this ``pair--rotate--orthonormalize'' procedure yields an equilibrated basis that underpins their explicit construction of an optimal anti-symmetric perturbation attaining the maximal spectral gap \cite{lelievre2013optimal}. 

Finally, asymptotic optimality does not necessarily translate into the best finite-time performance, since convergence bounds typically include a multiplicative prefactor in addition to the exponential rate. In non-reversible dynamics, a larger asymptotic rate may be accompanied by transient amplification due to non-normality, which is reflected in this prefactor $C$. Hence, the practical speed-up at the moderate times relevant for sampling and phase-transition phenomena depends on both the optimal exponent from Proposition~\ref{prop:maxgap} and the magnitude of $C$ \cite{lelievre2013optimal}. Optimizing $C$ while achieving the maximal asymptotic rate remains open; therefore, we refer to our optimized anti-symmetric perturbation $\mathbf{Q}$ as Leli\`evre's exponentially-optimal.

\section{Phase Transitions in the Generative Process with Non-Reversible Drift}

This section presents our main theoretical contribution and numerical experiments on a Gaussian mixture model, providing a detailed analysis of how the generalized drift matrix $\mathbf{A}$ influences the critical phase transitions of the diffusion process. We focus on the two key dynamical events identified by Biroli et al.: speciation and collapse \cite{biroli2024dynamical}. For each transition, our derivations begin with the most general, non-reversible case and subsequently demonstrate how our results unify and extend previous work by simplifying to known criteria in more restricted scenarios.

\subsection{The Speciation Transition}

The first critical event in the generative process is the \emph{speciation transition}, which marks the time $t_S$ when the model's trajectory commits to a specific class of data, such as cats versus dogs \cite{biroli2024dynamical}. Intuitively, this commitment happens when the latent structure of the data becomes detectable against the diffusion-induced noise background. In our formulation, this detectability is naturally quantified through second-order statistics, because both the forward OU dynamics and the mixture geometry propagate through covariances. This motivates a stability viewpoint: speciation corresponds to the moment when the log-density changes curvature in a way that makes symmetry breaking favorable. Consequently, $t_S$ can be defined as a precise instability time for the evolving distribution, and Landau theory provides the appropriate tool to derive its criterion.

\subsubsection{General Criterion from Landau Theory}

We characterize speciation as the onset of a geometric instability in the effective potential $-\ln P_t(\mathbf{x})$, namely when its local curvature loses positive definiteness. This yields a compact, operational criterion in terms of a single eigenvalue crossing:
\begin{align}
    \lambda_{\min}\!\left(\widetilde{\mathbf{M}}(t_S)\right)=0,
    \qquad
    \widetilde{\mathbf{M}}(t)
    =\mathbf{\Sigma}_{\mathrm{sto}}(t)-e^{-\mathbf{A}t}\mathbf{\Sigma}_{\mathrm{B}}(e^{-\mathbf{A}t})^\top,
    \label{eq:ts_general}
\end{align}
where $\mathbf{\Sigma}_{\mathrm{sto}}(t)$ is the accumulated stochastic covariance generated by the dynamics up to time $t$, and the second term propagates the symmetry-breaking content of the initial data forward under the linear flow $e^{-\mathbf{A}t}$. In other words, \autoref{eq:ts_general} directly compares the \emph{noise-built} covariance with the \emph{remaining signal} covariance at time $t$.
The condition in \autoref{eq:ts_general} matches the instability of the quadratic form derived in Appendix~\ref{app:landau}.
The two formulations differ only by a congruence transformation with the positive-definite factor $\mathbf{\Sigma}_{\mathrm{sto}}^{-1/2}(t)$, so the smallest eigenvalue crosses zero at the same time in either representation.

The matrix $\mathbf{\Sigma}_{\mathrm{B}}$ extracts the component of the data covariance that drives the unimodal-to-multimodal bifurcation, as opposed to isotropic broadening. Writing
\begin{align}
    \mathbf{\Sigma}(t_{0})
    =\langle \mathbf{a}\mathbf{a}^\top\rangle-\langle \mathbf{a}\rangle\langle \mathbf{a}\rangle^\top,
\end{align}
where $\mathbf{a}$ belongs to the Gaussian mixture, representing the data point. We set $\mathbf{\Sigma}_{\mathrm{B}}$ to the symmetry-breaking part of $\mathbf{\Sigma}(t_{0})$. For a two-component Gaussian mixture model, this decomposition takes the simple form
\begin{align}
    \mathbf{\Sigma}_{\mathrm{B}}=\mathbf{\Sigma}(t_{0})-\sigma^2\mathbf{I}=\mathbf{m}\mathbf{m}^\top,
\end{align}
so speciation is controlled by the projected separation mode $\mathbf{m}\mathbf{m}^\top$ rather than the isotropic intra-cluster variance $\sigma^2\mathbf{I}$. A detailed Landau expansion and the construction of $\mathbf{\Sigma}_{\mathrm{B}}$ are deferred to \autoref{app:landau}.

\subsubsection{Analysis in Special Cases}

Our general criterion provides a unified view, and it simplifies gracefully to known closed-form expressions under additional structure. The key simplification occurs when the relevant matrices share eigenvectors, so that the multidimensional instability reduces to a scalar condition along the dominant symmetry-breaking direction. This reduction is not merely technical: it clarifies which direction in data space controls the first onset of mode separation. In what follows we state the simultaneously diagonalizable case and then recover the standard isotropic result as a further specialization. 

\emph{Orthogonally Diagonalizable}: Assume that the drift $\mathbf{A}$ and the effective symmetry-breaking signal $\mathbf{\Sigma}_{\mathrm{B}}$ are \emph{simultaneously orthogonally diagonalizable}, i.e., there exists
an orthogonal matrix $\mathbf{P}$ such that
$\mathbf{P}^\top \mathbf{A}\mathbf{P}=\mathrm{diag}(d_1,\dots,d_d)$ with $d_i>0$ and
$\mathbf{P}^\top \mathbf{\Sigma}_{\mathrm{B}}\mathbf{P}=\mathrm{diag}(c_1,\dots,c_d)$.
For instance, this holds when $\mathbf{A}$ is symmetric and shares an orthonormal eigenbasis with $\mathbf{\Sigma}_{\mathrm{B}}$.
Let $\Lambda=\max_i c_i$ be the principal eigenvalue of $\mathbf{\Sigma}_{\mathrm{B}}$  with corresponding eigenvector
$\mathbf{v}_\Lambda$, and denote by $d_\Lambda$ the eigenvalue of $\mathbf{A}$ along the same direction,
$\mathbf{A}\mathbf{v}_\Lambda=d_\Lambda \mathbf{v}_\Lambda$.
Then the general instability criterion \autoref{eq:ts_general} reduces to a scalar equation along $\mathbf{v}_\Lambda$,
yielding the closed-form speciation time
\begin{align}
t_S=\frac{\ln(1+\Lambda d_\Lambda)}{2d_\Lambda}.
\label{eq:ts_commuting}
\end{align}
A detailed derivation is provided in Appendix~\ref{app:ts_commuting_derivation}. 

\emph{Isotropic Case ($\mathbf{A} = \mathbf{I}$):} For the standard isotropic model studied by Biroli et al. \cite{biroli2024dynamical}, the drift eigenvalue is simply $d_\Lambda=1$. Substituting this specialization into \autoref{eq:ts_commuting} immediately yields the isotropic expression for $t_S$ in terms of the signal strength $\Lambda$. Since the signal strength $\Lambda$ typically scales with dimension $d$, the solution further simplifies to their original result:
\begin{align}
    t_S = \frac{\ln(1 + \Lambda)}{2} \approx \frac{1}{2}\ln(\Lambda).
\end{align}

\subsection{The Collapse Transition}

The second critical event is the \emph{collapse transition}, which occurs at time $t_C$ when the generative process ceases to generalize and instead ``collapses'' onto specific data points from the training set, entering a regime of pure memorization \cite{biroli2024dynamical}. Conceptually, this transition marks a qualitative change in the effective support of the distribution produced by the reverse dynamics: instead of covering a continuum of plausible variations, the mass concentrates around discrete training exemplars. This distinction is crucial because it separates a model that can generate novel samples from one that merely reproduces its training data. In our analysis, the collapse time will be characterized by an information-theoretic criterion that depends on the entropy of the evolving distribution and the number of stored patterns. 

\subsubsection{General Criterion from an Entropic Volume Argument}

A convenient physical intuition for the collapse transition comes from an entropic volume argument in high dimensions, where entropy directly controls the effective size of typical sets. The effective volume $V$ occupied by a probability distribution $P(\mathbf{x})$ is related to its Shannon entropy $S$ by $V \approx e^S$ \cite{biroli2024dynamical}, so comparing volumes becomes equivalent to comparing entropies. A collapse occurs when the effective volume of the true generative distribution, $V_{\mathrm{true}}(t) = e^{d s(t)}$, becomes comparable to the minimal volume required to store the $n$ training samples as distinct, separated Gaussian lumps, $V_{\mathrm{sep}}(t) = n \cdot V_G(t) = n e^{d s_G(t)}$ \cite{biroli2024dynamical}. Equating these volumes yields a general criterion on the per-variable entropies:
\begin{align}
    s(t_C) = s^{\mathrm{sep}}(t_C) = \frac{\ln n}{d} + s_G(t_C), \label{eq:tc_general}
\end{align}
where $s(t)$ is the entropy density of the true data distribution $P_t(\mathbf{x})$, and $s_G(t)$ is the entropy density of a single Gaussian component with covariance $\Sig_{\mathrm{sto}}(t)$:
\begin{align}
    s_G(t) = \frac{1}{2}\log(2\pi e) + \frac{1}{2d}\log(\det \Sig_{\mathrm{sto}}(t)).
\end{align}
Interpreting $t$ as the forward noising time, the criterion separates regimes in terms of the reverse process initialized at noise level $t$: for $t<t_C$ the reverse dynamics is dominated by memorization around individual training exemplars, whereas for $t>t_C$ it supports a generalized regime with nontrivial variability \cite{biroli2024dynamical}.

\subsubsection{Analytical Insight from the Random Energy Model}

To obtain analytical insight into the collapse phenomenon beyond the general entropic criterion, we focus on a special case where the drift matrix is symmetric and diagonal, $\mathbf{A} = \mathbf{U}$. In this setting, the forward dynamics admits a controlled mapping to a Random Energy Model (REM), which is a classic statistical-physics framework for glassy systems \cite{biroli2024dynamical}. The REM viewpoint makes the mechanism explicit by separating the contribution of the ``correct'' data point ($Z'_1$) from the collective background of all other data points ($Z'_{2...n}$), so collapse corresponds to the point where the background ceases to be negligible. By analyzing the competition between these terms, one can compute the free-energy densities and thereby obtain an implicit equation for $t_C$ that is fully determined by the microscopic parameters of the model. The REM reduction provides a concrete route from the qualitative entropy picture to an explicit solvable criterion for the collapse time.

In particular, the final condition emerges from equating the free energy densities of the signal and the background, leading to the criterion $\alpha + g_t(1)|_{t_C} = -1/2$, where $\alpha = (\ln n)/d$ and $g_t(1)$ is the free energy density of the background system. Translating this equality into the OU parameters yields the implicit equation for the collapse time:
\begin{align}
    2\alpha + 1 = \frac{1}{d} \sum_{j=1}^{d} \left[ \ln\left(1 + 2\sigma_0^2 \frac{u_j e^{-2u_j t_C}}{1-e^{-2u_j t_C}}\right) + \frac{1}{1 + 2\sigma_0^2 \frac{u_j e^{-2u_j t_C}}{1-e^{-2u_j t_C}}} \right], \label{eq:tc_rem}
\end{align}
where $u_j$ are the diagonal elements of $\mathbf{U}$ and $\sigma_0^2$ characterizes the variance of the data. This expression makes the dependence on anisotropy explicit through the full spectrum $\{u_j\}_{j=1}^d$, while the dataset size enters only through $\alpha=(\ln n)/d$. As a result, the macroscopic collapse time is directly tied to microscopic data statistics and the anisotropic learning dynamics encoded by $\mathbf{U}$. 

\subsection{Analysis of Gaussian Mixture Models}

We use the Gaussian mixture model to study the backward dynamics in the large $d$ and $n$ limit. In this setting, the initial law $P_0(\mathbf{a})$ is the superposition of two Gaussian clusters of equal weight, which we take for simplicity with means $\mathbf{m}$ and the same variance $\sigma^2\mathbf{I}$. This choice provides a minimal yet nontrivial multimodal structure in which the speciation and collapse transitions can be probed cleanly. Without loss of generality, and also to simplify the calculation, we will use the diagonal potential energy matrix throughout the following discussion. 
It is important to note that the covariance matrix of this Gaussian mixture is given by:
\begin{align}
     \Sig_{\mathrm{data}} = \mathbf{C_\mathrm{total}}=  \sigma^2\mathbf{I} + \mathbf{m}\mathbf{m}^\top.
\end{align}
This decomposition separates the isotropic within-cluster variance from the rank-one between-cluster signal carried by $\mathbf{m}\mathbf{m}^\top$, which is the symmetry-breaking component relevant for speciation. As a result, the same object naturally enters multiple routes for estimating or predicting $t_S$, depending on whether one works with covariances, order parameters, or stability criteria. In particular, it clarifies why different derivations can be consistent while using apparently different starting points, since they ultimately track the same rank-one signal against the stochastic background. 

\textit{Speciation time.} We estimate the speciation time $t_S$ using the trajectory-cloning protocol of Ref.~\cite{biroli2024dynamical}.
A backward trajectory is initialized at $t_f\gg 1$ from $\mathbf{x}_f\sim\mathcal{N}(0,\Sig_s)$ with $\Sig_s=\mathbf{U}^{-1}$.
At each forward time $t$, two independent reverse trajectories are cloned from the same state $\mathbf{y}$, and we record whether they end in the same mixture component at time $0$.
The corresponding probability is denoted by $\phi(t)$.

For a symmetric two-component Gaussian mixture, the forward-time density at time $t$ can be written as
\begin{align}
P_t(\mathbf{y})
=\frac{1}{2}G\!\left(\mathbf{y},\boldsymbol{\mu}(t),\Sig_G(t)\right)
+\frac{1}{2}G\!\left(\mathbf{y},-\boldsymbol{\mu}(t),\Sig_G(t)\right),
\end{align}
where $\boldsymbol{\mu}(t)=e^{-\mathbf{A}t}\mathbf{m}$ and $\Sig_G(t)$ is the within-component covariance after forward noising.
Starting from an initial within-component covariance $\sigma^2\mathbf{I}$, it takes the form
\begin{align}
\Sig_G(t)=\Sig_{\mathrm{sto}}(t)+\sigma^2\,e^{-\mathbf{A}t}e^{-\mathbf{A}^\top t}.
\end{align}
Here $G(\mathbf{y},\boldsymbol{\mu},\Sig)$ denotes the $d$-dimensional Gaussian density with mean $\boldsymbol{\mu}$ and covariance $\Sig$.

Bayes' rule gives the posterior weights of the two components conditioned on $\mathbf{y}$,
\begin{align}
\phi(t)
&=\frac{1}{2}\int d\mathbf{y}\,
\frac{
G\!\left(\mathbf{y},\boldsymbol{\mu}(t),\Sig_G(t)\right)^2
+
G\!\left(\mathbf{y},-\boldsymbol{\mu}(t),\Sig_G(t)\right)^2
}{
G\!\left(\mathbf{y},\boldsymbol{\mu}(t),\Sig_G(t)\right)
+
G\!\left(\mathbf{y},-\boldsymbol{\mu}(t),\Sig_G(t)\right)
}
\nonumber\\
&=1-\int d\mathbf{y}\,
\frac{
G\!\left(\mathbf{y},\boldsymbol{\mu}(t),\Sig_G(t)\right)\,
G\!\left(\mathbf{y},-\boldsymbol{\mu}(t),\Sig_G(t)\right)
}{
G\!\left(\mathbf{y},\boldsymbol{\mu}(t),\Sig_G(t)\right)
+
G\!\left(\mathbf{y},-\boldsymbol{\mu}(t),\Sig_G(t)\right)
},
\label{eq:phi_identity}
\end{align}
To simplify the integrand in \autoref{eq:phi_identity}, write the Gaussian density as
\[
G(\mathbf{y},\boldsymbol{\mu},\Sig_G)=\frac{1}{\sqrt{(2\pi)^d\det\Sig_G}}
\exp\!\left(-\frac{1}{2}(\mathbf{y}-\boldsymbol{\mu})^\top \Sig_G^{-1}(\mathbf{y}-\boldsymbol{\mu})\right).
\]
With $\boldsymbol{\mu}(t)=e^{-\mathbf{A}t}\mathbf{m}$ and $\Sig_G(t)$ the within-component covariance at time $t$, one obtains:
\begin{align}
\phi(t)
= 1 - \frac{1}{2}\exp\!\left(-\frac{1}{2}\boldsymbol{\mu}(t)^\top\Sig_G(t)^{-1}\boldsymbol{\mu}(t)\right)
\int \frac{d\mathbf{y}}{\sqrt{(2\pi)^d\det\Sig_G(t)}}
\exp\!\left(-\frac{1}{2}\mathbf{y}^\top\Sig_G(t)^{-1}\mathbf{y}\right)
\frac{1}{\cosh\!\left(\mathbf{y}^\top\Sig_G(t)^{-1}\boldsymbol{\mu}(t)\right)}.
\label{eq:phi_cosh_form}
\end{align}
For numerical evaluation it is convenient to whiten the Gaussian measure by the change of variables
$\mathbf{z}=\Sig_G(t)^{-1/2}\mathbf{y}$, which turns the Gaussian weight into $\exp(-\|\mathbf{z}\|^2/2)$.

\begin{figure}[ht!] 
  \centering 
    \includegraphics[width=0.9\textwidth]{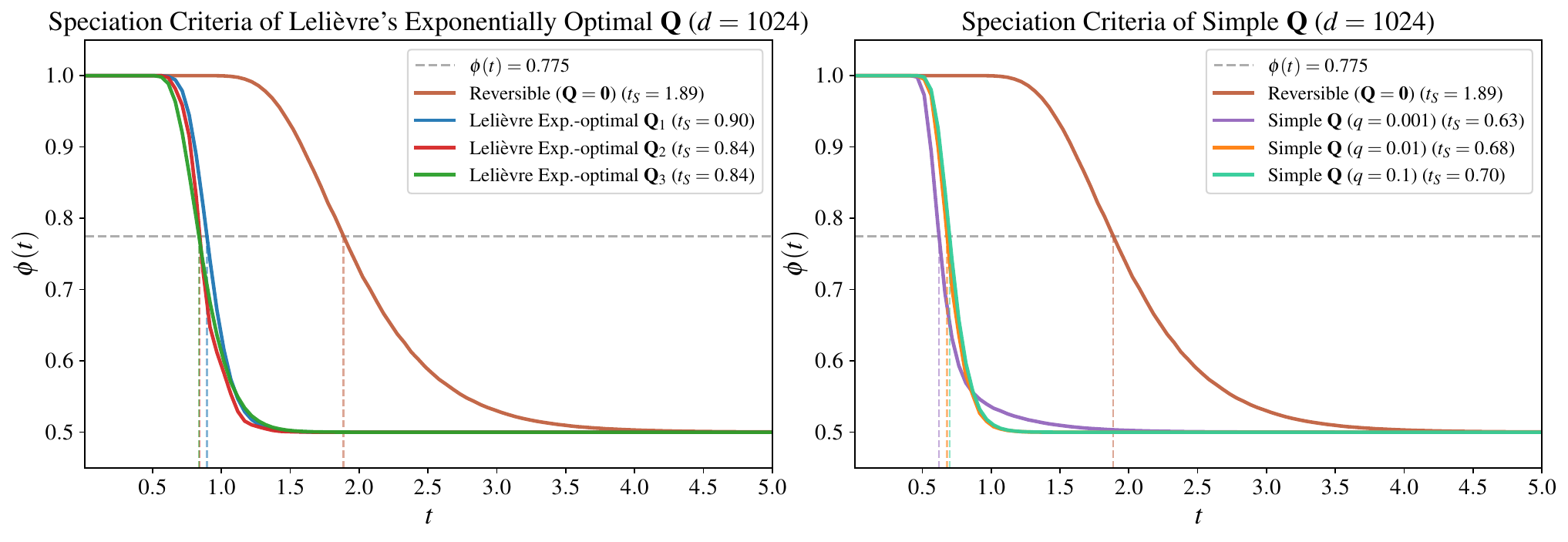}
  \caption{Acceleration of speciation by non-reversible drift in Gaussian mixture. Left: Dynamics under Leli\`evre's exponentially optimal drift. The labels $\mathbf{Q}_1$, $\mathbf{Q}_2$, and $\mathbf{Q}_3$ correspond to different choices of the auxiliary spectrum used in the construction algorithm. $\mathbf{Q}_1$ uses a standard linear spectrum with $\lambda \in [1, d]$, $\mathbf{Q}_2$ uses a shifted high-frequency spectrum with $\lambda \in [d+1, 2d]$, and $\mathbf{Q}_3$ uses a geometric spectrum. Right: Dynamics under a simple $\mathbf{Q}$ strategy defined as a dense anti-symmetric matrix where all upper-triangular elements are set to a constant magnitude $q$, specifically $Q_{ij} = q$ for $i < j$. The theoretical speciation times $t_S$ in legends are computed by solving $\lambda_{\min}(\widetilde{\mathbf{M}}(t_S))=0$. Notably, the simple strategy achieves faster speciation than the ``exponentially optimal'' designs. This arises because Leli\`evre's optimality applies to the asymptotic decay rate when $t \to \infty$, whereas the speciation event occurs at short times where transient non-normal effects dominate the dynamics.}
  \label{fig:phi_t_comparison}
\end{figure}

In the isotropic reversible case studied in \cite{biroli2024dynamical}, $\mathbf{A}=\mathbf{I}$ and $\Sig_G(t)=\Gamma_t\mathbf{I}$, so all directions orthogonal to $\mathbf{m}$ factorize and the expression reduces to the one-dimensional integral reported there.
For the non-reversible drifts used in Fig.~\ref{fig:phi_t_comparison}, $\Sig_G(t)$ is generally non-diagonal, and $\phi(t)$ is therefore measured directly from cloned simulations.

\begin{figure} 
  \centering 
    \includegraphics[width=0.9\textwidth]{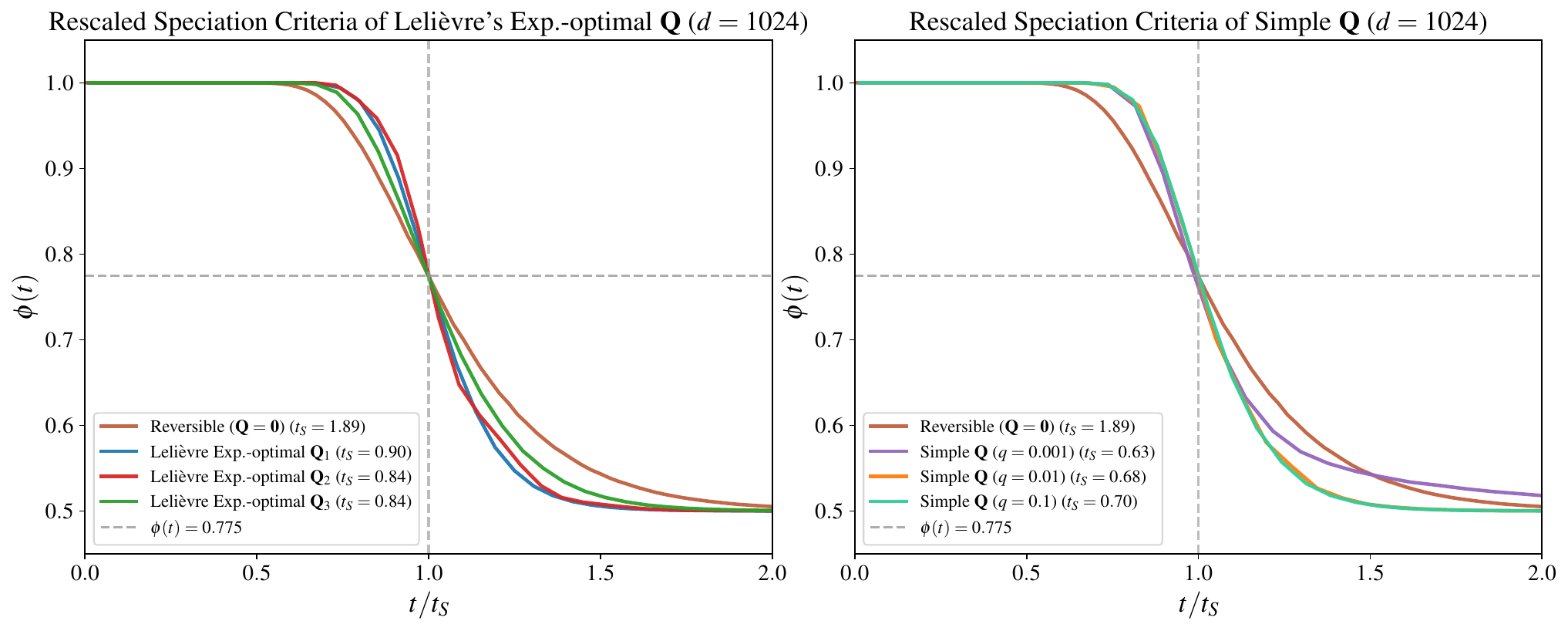}
  \caption{Validation of the theoretical scaling for speciation dynamics. The probability $\phi(t)$ is plotted against the rescaled time $t/t_S$ using the same experimental configuration and drift matrices defined in Figure 1. Left: Results for Leli\`evre's exponentially optimal drift. Right: Results for the simple $\mathbf{Q}$ strategy. The theoretical $t_S$ matches perfectly with the numerical results obtained from the Gaussian mixture simulations.
}
  \label{fig:phi_t_normalized}
\end{figure}

While Fig.~\ref{fig:phi_t_comparison} demonstrates that non-reversible drifts significantly accelerate the absolute speed of speciation, we further test whether our theoretical derivation for the speciation time $t_S$ captures the transition onset consistently across these diverse regimes. 
Following standard finite-size scaling, we rescale the time axis by the predicted $t_S$ for each strategy, which removes strategy-dependent time units and isolates the instability timescale. As shown in Fig.~\ref{fig:phi_t_normalized}, this normalization brings the rapid change of $\phi(t)$ into a narrow window around $t/t_S=1$, demonstrating that the transition is triggered at a comparable rescaled time despite markedly different transient behaviors in physical time.

\textit{Collapse time.} We next turn to the collapse transition and assess whether non-reversibility affects the entropy-based criterion for $t_C$ in the same way it affects speciation. Our numerical experiments show that when the potential matrix $\mathbf{U}$ is diagonal, the value of $\det \Sig_{\mathrm{sto}}(t)$ at a fixed time is not affected by the choice of $\mathbf{Q}$. For a non-diagonal potential matrix, $\det \Sig_{\mathrm{sto}}(t)$ at a fixed time is affected by $\mathbf{Q}$, but this does not change the zero crossing of the excess entropy density. This indicates that, although $\mathbf{Q}$ can reshape transient covariance evolution in more general anisotropic settings, the collapse onset remains robust as predicted by our theoretical invariance arguments. 

\begin{figure}[h!] 
  \centering 
    \includegraphics[width=0.9\linewidth]{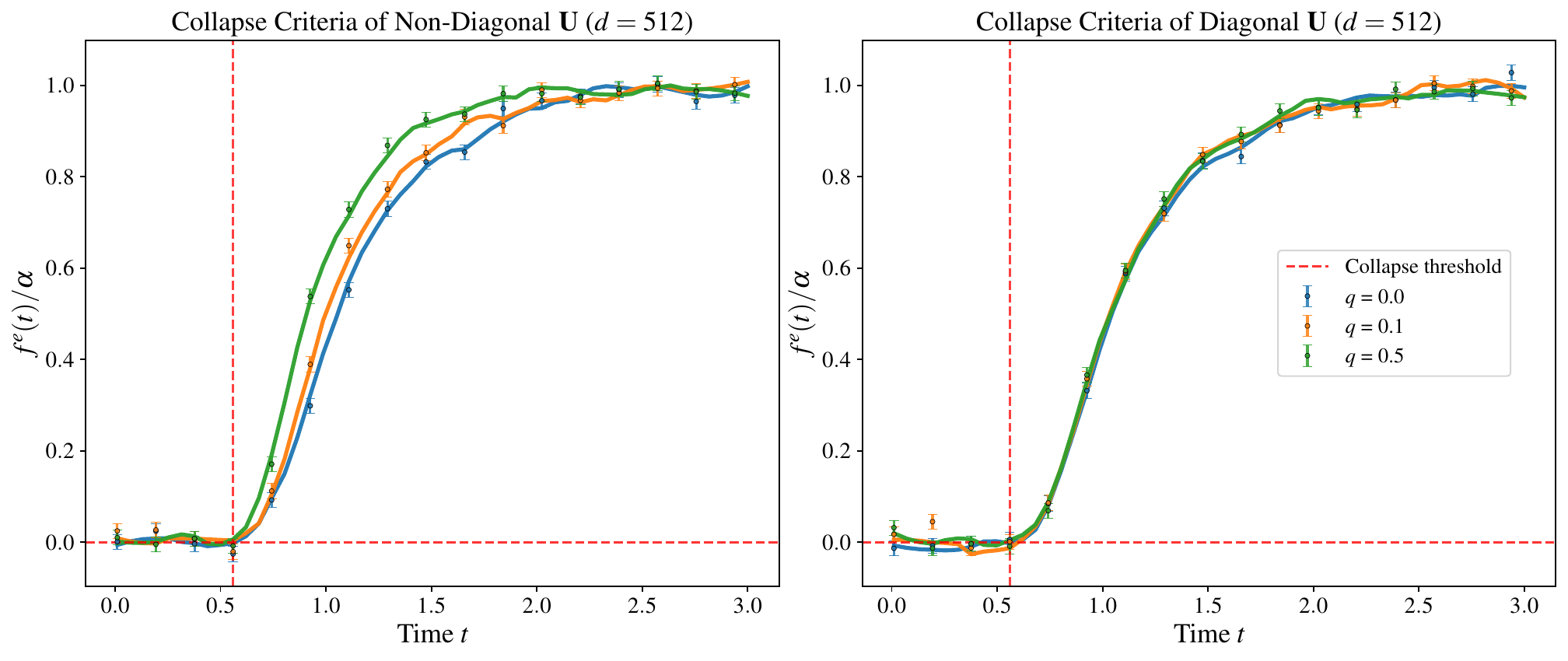}
  \caption{Invariance of the collapse time to non-reversible perturbations. The normalized excess entropy density is plotted against time for different values of $\mathbf{Q}$. The collapse time $t_C$ is defined as the time at which the curves lift off from zero. Left: For non-diagonal $\mathbf{U}$, while the shapes of the transition curves differ, their onset point remains the same regardless of the value of $\mathbf{Q}$, validating that the collapse time is robust to non-reversible perturbations. Right:For diagonal $\mathbf{U}$, the entropy curves for different $\mathbf{Q}$ values are indistinguishable, showing that both the collapse time and subsequent dynamics are independent of the non-reversible component.}
\label{fig:collapse_exp}
\end{figure}

\section{Discussion}

Our results highlight a clean separation between the roles of non-reversible currents and entropic contraction in diffusion dynamics. By introducing an anti-symmetric component $\mathbf{Q}$ into the linear drift $\mathbf{A}=(\mathbf{I}+\mathbf{Q})\mathbf{U}$, one can reshape probability currents and increase the effective relaxation rate while keeping the invariant measure fixed. Within this class, exponentially optimal constructions provide an operational route to equalize decay rates across directions and thereby accelerate long-time mixing. In the statistical-physics picture, this acceleration translates into an earlier onset of the speciation transition, since the symmetry-breaking instability is reached sooner in absolute time. Our controlled violations of detailed balance offer a principled and practically usable knob to speed up mode separation without changing the stationary target.

A complementary message is that the collapse transition is controlled by a purely entropic phase-space contraction mechanism and is therefore robust to non-reversible perturbations. Because the relevant volume contraction rate depends only on $\Tr(\mathbf{A})=\Tr(\mathbf{U})$, adding $\mathbf{Q}$ does not shift the collapse time defined by the entropic-volume criterion. This robustness is consistent with recent empirical observations that collapse timing can remain stable under changes of coupling structure, while still allowing transient trajectories and currents to differ. In this sense, non-reversible control can improve practical sampling and relaxation efficiency without pulling the system closer to the memorization-dominated regime. 

This perspective also clarifies how our results relate to the synchronization-gap phenomenology reported in \cite{albrychiewicz2026dynamicalregimesmultimodaldiffusion}. Their gap reflects a coupling-induced separation of relaxation time scales across different channels of the coupled system. Our intervention acts at a different level: non-reversible currents redistribute relaxation pathways within a fixed invariant measure, so they can reduce the absolute time needed to reach the speciation instability without shifting the entropic collapse threshold. Taken together, the two viewpoints suggest a useful diagnostic for practice.

Several extensions remain, which suggest a future roadmap. First, our analysis focuses on linear forward processes and idealized high-dimensional data models, so extending the same control principles to nonlinear drifts, learned score networks, and realistic datasets is an important next step. Second, it will be valuable to connect the continuous-time exponentially optimal construction more directly to implementable discrete-time samplers and to quantify trade-offs under fixed compute budgets, discretization error, and numerical stability. Third, beyond optimizing the asymptotic rate, understanding and controlling finite-time prefactors, such as the constant $C$ arising from non-normal transient effects, may be crucial for practical gains at the time scales relevant for sampling and phase-transition phenomena.

\section*{Acknowledgments} 
This work is supported by Projects 12322501, 12575035 of the National Natural Science Foundation of China, and 2026NSFSCZY0124 of the Natural Science Foundation of Sichuan Province. 
The computational work is supported by the Center for HPC, University of Electronic Science and Technology of China. The code of this study is available from the corresponding author upon reasonable request.

\section*{Author contributions}
Y.T. had the original idea for this work. H.L. performed the study and wrote the manuscript.

\section*{Competing interests} 

The authors declare no competing interests.

\bibliography{refs_final.bib}
\newpage

\appendix

\section{Derivation of the Speciation Criterion}
\label{app:landau}

\subsection{Landau Expansion}
This appendix provides a detailed derivation of the speciation criterion presented in \autoref{eq:ts_general}. We start from the Landau-type expansion of the log-probability of the empirical distribution at time $t$. The distribution is given by:
\begin{align}
    P_t(\mathbf{x}) = \frac{1}{\sqrt{\det(2\pi\mathbf{\Sigma}_{\mathrm{sto}}(t))}} \exp\left(-\frac{1}{2} \mathbf{x}^\top \mathbf{\Sigma}_{\mathrm{sto}}^{-1}(t) \mathbf{x} + g(\mathbf{x})\right),
\end{align}
where the generating function $g(\mathbf{x})$ is defined as:
\begin{align}
    g(\mathbf{x}) = \ln \int d\mathbf{a} \, P_0(\mathbf{a}) \exp\left(-\frac{1}{2} (e^{-\mathbf{A} t} \mathbf{a})^\top \mathbf{\Sigma}_{\mathrm{sto}}^{-1}(t) (e^{-\mathbf{A} t} \mathbf{a}) + \mathbf{x}^\top \mathbf{\Sigma}_{\mathrm{sto}}^{-1}(t) e^{-\mathbf{A} t} \mathbf{a}\right).
\end{align}
When $\|e^{-\mathbf{A}t}\|$ is small, the integral admits a Taylor expansion in powers of $e^{-\mathbf{A}t}$. Let $\Phi(\mathbf{a}; \mathbf{x}, t) = -\frac{1}{2}\,\mathbf{a}^\top\,(e^{-\mathbf{A}t})^\top\,\mathbf{\Sigma}_{\mathrm{sto}}^{-1}\,e^{-\mathbf{A}t}\,\mathbf{a} + \mathbf{x}^\top\,\mathbf{\Sigma}_{\mathrm{sto}}^{-1}\,e^{-\mathbf{A}t}\,\mathbf{a}$. We expand $\exp(\Phi) \approx 1 + \Phi + \frac{1}{2}\Phi^2 + \dots$.

We then integrate term by term with respect to the data distribution $P_0(\mathbf{a})$, where $\langle \cdot \rangle$ denotes expectation.
The term independent of $\mathbf{x}$ contributes:
\begin{align}
    \langle \mathrm{const} \rangle = 1 - \frac{1}{2} \Tr\left( (e^{-\mathbf{A}t})^\top \mathbf{\Sigma}_{\mathrm{sto}}^{-1} e^{-\mathbf{A}t} \langle \mathbf{a}\mathbf{a}^\top \rangle \right).
\end{align}
The term linear in $\mathbf{x}$ contributes:
\begin{align}
    \langle \mathrm{linear} \rangle = \mathbf{x}^\top \mathbf{\Sigma}_{\mathrm{sto}}^{-1} e^{-\mathbf{A}t} \langle \mathbf{a} \rangle.
\end{align}
The term quadratic in $\mathbf{x}$ contributes:
\begin{align}
    \langle \mathrm{quadratic} \rangle = \frac{1}{2} \mathbf{x}^\top \mathbf{\Sigma}_{\mathrm{sto}}^{-1} e^{-\mathbf{A}t} \langle \mathbf{a}\mathbf{a}^\top \rangle (e^{-\mathbf{A}t})^\top \mathbf{\Sigma}_{\mathrm{sto}}^{-1} \mathbf{x}.
\end{align}
Substituting these into the expression for $g(\mathbf{x})$ and using the expansion $\ln(1+z) \approx z - \frac{1}{2}z^2$, we collect all terms up to second order in $\mathbf{x}$:
\begin{align}
    g(\mathbf{x}) \approx \mathrm{Const} + \mathbf{x}^\top \mathbf{\Sigma}_{\mathrm{sto}}^{-1} e^{-\mathbf{A}t} \langle \mathbf{a} \rangle + \frac{1}{2} \mathbf{x}^\top \mathbf{\Sigma}_{\mathrm{sto}}^{-1} e^{-\mathbf{A}t} \langle \mathbf{a}\mathbf{a}^\top \rangle (e^{-\mathbf{A}t})^\top \mathbf{\Sigma}_{\mathrm{sto}}^{-1} \mathbf{x} - \frac{1}{2} \left( \mathbf{x}^\top \mathbf{\Sigma}_{\mathrm{sto}}^{-1} e^{-\mathbf{A}t} \langle \mathbf{a} \rangle \right)^2.
\end{align}
Combining this with the quadratic term from the Gaussian prefactor in $P_t(\mathbf{x})$, the full quadratic form in the exponent is $-\frac{1}{2}\mathbf{x}^\top\mathbf{M}(t)\mathbf{x}$. Here, the total curvature matrix $\mathbf{M}(t)$ is determined by the full covariance of the data, denoted as $\mathbf{\Sigma}(t_{0}) = \langle \mathbf{a}\mathbf{a}^\top \rangle - \langle \mathbf{a} \rangle\langle \mathbf{a} \rangle^\top$:
\begin{align}
    \mathbf{M}(t) &= \mathbf{\Sigma}_{\mathrm{sto}}^{-1}(t) - \mathbf{\Sigma}_{\mathrm{sto}}^{-1}(t) e^{-\mathbf{A}t} \left( \langle \mathbf{a}\mathbf{a}^\top \rangle - \langle \mathbf{a} \rangle\langle \mathbf{a} \rangle^\top \right) (e^{-\mathbf{A}t})^\top \mathbf{\Sigma}_{\mathrm{sto}}^{-1}(t) \\
    &= \mathbf{\Sigma}_{\mathrm{sto}}^{-1}(t) - \mathbf{\Sigma}_{\mathrm{sto}}^{-1}(t) e^{-\mathbf{A}t} \mathbf{\Sigma}(t_{0}) (e^{-\mathbf{A}t})^\top \mathbf{\Sigma}_{\mathrm{sto}}^{-1}(t).
\end{align}

Physically, the speciation transition corresponds to a spontaneous symmetry breaking event where the distribution bifurcates from unimodal to multimodal. It is therefore crucial to distinguish between two distinct types of geometric instabilities contained within $\mathbf{\Sigma}(t_{0})$:
\begin{enumerate}
    \item Isotropic Fluctuations: Contributions from the isotropic intra-class variance ($\sigma^2 \mathbf{I}$) represent the background Gaussian fluctuations of the system. 
    In the language of Landau theory, these modes correspond to a renormalization of the effective temperature, equivalently the distribution width. An instability in this sector leads to a broadening of the probability density but preserves its unimodal topology, and thus does not drive the phase transition.
    \item Order Parameter Instability: Contributions from the cluster separation ($\mathbf{m}\mathbf{m}^\top$) project onto the longitudinal direction of the potential landscape. This term represents the critical mode associated with the order parameter. An instability here drives the topological transition from a single mode to distinct modes, marking the onset of speciation.    
\end{enumerate}
To precisely identify the speciation time $t_S$, we must isolate the order parameter driving the bifurcation. In the presence of non-reversible drifts ($\mathbf{Q}\neq 0$) where modes can couple, we define the effective symmetry-breaking signal $\mathbf{\Sigma}_{\mathrm{B}}$ as the projection of the data covariance onto the order-parameter subspace. For the Gaussian mixture model, this reads:
\begin{align}
    \mathbf{\Sigma}_{\mathrm{B}} = \mathbf{\Sigma}(t_{0}) - \sigma^2 \mathbf{I} = \mathbf{m}\mathbf{m}^\top.
\end{align}
The transition occurs when the curvature destabilizes specifically due to this signal. This yields the condition on the effective stability matrix $\widetilde{\mathbf{M}}(t)$:
\begin{align}
    \lambda_{\min}(\widetilde{\mathbf{M}}(t_S)) = 0, \quad \mathrm{with} \quad \widetilde{\mathbf{M}}(t) = \mathbf{\Sigma}_{\mathrm{sto}}(t) - e^{-\mathbf{A}t} \mathbf{\Sigma}_{\mathrm{B}} (e^{-\mathbf{A}t})^\top. \label{eq:ts_general}
\end{align}
This completes the derivation of the general criterion for the speciation time $t_S$.

\subsection{Closed-form Solution in the Simultaneously Diagonalizable Case}
\label{app:ts_commuting_derivation}

Starting from the general criterion (see \autoref{eq:ts_general}),
\begin{align}
\lambda_{\min}\!\left(\widetilde{\mathbf{M}}(t)\right)=0,
\qquad
\widetilde{\mathbf{M}}(t)=\Sigma_{\mathrm{sto}}(t)-e^{-\mathbf{A}t}\mathbf{\Sigma}_{\mathrm{B}} e^{-\mathbf{A}^\top t},
\label{eq:app_ts_general}
\end{align}
we assume that $\mathbf{A}$ and the symmetry-breaking signal $\mathbf{\Sigma}_{\mathrm{B}}$ are simultaneously orthogonally diagonalizable: there exists an
orthogonal matrix $\mathbf{P}$ such that
\begin{align}
\mathbf{P}^\top \mathbf{A}\mathbf{P}=\mathrm{diag}(d_1,\dots,d_d),\qquad
\mathbf{P}^\top \mathbf{\Sigma}_{\mathrm{B}}\mathbf{P}=\mathrm{diag}(c_1,\dots,c_d),
\qquad d_i>0.
\label{eq:app_simul_diag}
\end{align}
Define the rotated coordinates $\mathbf{x}'=\mathbf{P}^\top\mathbf{x}$. In this basis,
\begin{align}
e^{-\mathbf{A}t}=\mathbf{P}\,\mathrm{diag}(e^{-d_1 t},\dots,e^{-d_d t})\,\mathbf{P}^\top,
\qquad
e^{-\mathbf{A}^\top t}=e^{-\mathbf{A}t},
\label{eq:app_expA}
\end{align}
and the stochastic covariance (using $\Sigma_{\mathrm{sto}}(t)=2\int_0^t e^{-\mathbf{A}s}e^{-\mathbf{A}^\top s}ds$)
becomes diagonal:
\begin{align}
\Sigma_{\mathrm{sto}}(t)
&=\mathbf{P}\,\mathrm{diag}\!\left(\frac{1-e^{-2d_1 t}}{d_1},\dots,\frac{1-e^{-2d_d t}}{d_d}\right)\mathbf{P}^\top.
\label{eq:app_Sigma_sto_diag}
\end{align}
Similarly, the rotated signal term becomes:
\begin{align}
e^{-\mathbf{A}t}\mathbf{\Sigma}_{\mathrm{B}} e^{-\mathbf{A}^\top t}
=\mathbf{P}\,\mathrm{diag}\!\left(c_1 e^{-2d_1 t},\dots,c_d e^{-2d_d t}\right)\mathbf{P}^\top.
\label{eq:app_signal_diag}
\end{align}
Therefore, in the common eigenbasis the matrix $\widetilde{\mathbf{M}}(t)$ is diagonal with entries
\begin{align}
\widetilde{m}_i(t)=\frac{1-e^{-2d_i t}}{d_i}-c_i e^{-2d_i t}.
\label{eq:app_mi}
\end{align}
The instability $\lambda_{\min}(\widetilde{\mathbf{M}}(t))=0$ occurs when at least one $\widetilde{m}_i(t)$ vanishes.
For the direction corresponding to the principal eigenvalue $\Lambda=\max_i c_i$, we denote the associated drift rate by
$d_\Lambda$. Setting \eqref{eq:app_mi} to zero gives
\begin{align}
\frac{1-e^{-2d_\Lambda t_S}}{d_\Lambda}=\Lambda e^{-2d_\Lambda t_S}
\quad\Longrightarrow\quad
e^{-2d_\Lambda t_S}=\frac{1}{1+\Lambda d_\Lambda},
\end{align}
hence
\begin{align}
t_S=\frac{\ln(1+\Lambda d_\Lambda)}{2d_\Lambda},
\end{align}
which is the closed-form expression stated in \autoref{eq:ts_commuting}.

\section{Random Energy Model Analysis for the Collapse Transition}
\label{app:rem}

\subsection{Proof of Volume Invariance under Non-Reversible Drift}
In this section, we explicitly compare our analytical results with the recent findings of Albrychiewicz et al.
\cite{albrychiewicz2026dynamicalregimesmultimodaldiffusion} to clarify the mathematical origin of the collapse time's
robustness.

Albrychiewicz et al. derived the collapse time $t_C$ based on the Random Energy Model, obtaining the implicit
condition (Eq.~3.23 in Ref.~\cite{albrychiewicz2026dynamicalregimesmultimodaldiffusion}):
\begin{align}
    \alpha = -\Lambda_{t_C}(1) - \frac{1}{2}, \label{eq:alb_collapse}
\end{align}
where $\Lambda_{t_C}(\beta)$ is the cumulant generating function of the energy distribution. They used this
formula to explore how $t_C$ behaves under different coupling structures and \textit{observed} that $t_C$ ``remains nearly
constant,'' providing helpful numerical guidance on the robustness of collapse. In the presentation of Ref.~\cite{albrychiewicz2026dynamicalregimesmultimodaldiffusion}, the emphasis is on solving \autoref{eq:alb_collapse} for specific coupling matrices, while a general mechanism explaining the robustness across non-reversible perturbations is not made explicit. Here we complement their analysis by showing that, under our entropic-volume definition of collapse (cf.~\autoref{eq:tc_general}), the invariance follows from a simple and general trace/volume-conservation argument that applies to arbitrary $\mathbf{Q}$.

Before focusing on the explicit solution for the diagonal case, we provide an analytical explanation that the
macroscopic collapse time is invariant under any non-reversible perturbation $\mathbf{Q}$, provided that the collapse
is defined by our entropic-volume criterion.

The key invariant we use is the \emph{linear contraction of the cloud of means} induced by the drift.
Indeed, conditioned on an initial point, the OU process has deterministic mean
$\boldsymbol{\mu}(t)=e^{-\mathbf{A}t}\boldsymbol{\mu}(0)$.
Hence the volume element transported by the linear map $\boldsymbol{\mu}(0)\mapsto \boldsymbol{\mu}(t)$ contracts as
\begin{align}
    \mathcal{V}_t
    = \mathcal{V}_0 \, \det\!\bigl(e^{-\mathbf{A}t}\bigr)
    = \mathcal{V}_0 \exp\!\left( - \int_0^t \Tr(\mathbf{A}) \, d\tau \right),
    \label{eq:mean_volume_contraction}
\end{align}
where we used $\det(e^{\mathbf{M}})=e^{\Tr(\mathbf{M})}$.
Within our entropic \emph{volume} definition of the collapse time (cf.~\autoref{eq:tc_general}),
the onset is controlled by the contraction of the mean cloud, and therefore by the leading exponential rate
$\Tr(\mathbf{A})$.

More precisely, in the standard entropic volume argument used in the main text (see \autoref{eq:tc_general}),
the mixture volume can be viewed as a product of a \emph{cloud-of-means} contribution and a \emph{single-lump} Gaussian
contribution. Since the separated reference volume is $V_{\mathrm{sep}}(t)=n\,V_G(t)$, the Gaussian lump factor
$V_G(t)$ cancels in the threshold equation, leaving the onset $t_C$ controlled by the contraction of the mean cloud,
whose leading exponential rate is set by $\Tr(\mathbf{A})$.

Using the property that the trace is invariant under transposition and
cyclic permutation, along with the definitions $\mathbf{U}^\top = \mathbf{U}$
and $\mathbf{Q}^\top = -\mathbf{Q}$, we derive:
\begin{align}
    \Tr(\mathbf{Q}\mathbf{U}) &= \Tr\left( (\mathbf{Q}\mathbf{U})^\top \right) \nonumber \\
    &= \Tr\left( \mathbf{U}^\top \mathbf{Q}^\top \right) \nonumber \\
    &= \Tr\left( \mathbf{U} (-\mathbf{Q}) \right) \nonumber \\
    &= -\Tr\left( \mathbf{U}\mathbf{Q} \right) \nonumber \\
    &= -\Tr\left( \mathbf{Q}\mathbf{U} \right).
\end{align}
Since $\Tr(\mathbf{Q}\mathbf{U}) = -\Tr(\mathbf{Q}\mathbf{U})$, it implies that $\Tr(\mathbf{Q}\mathbf{U}) = 0$.

This result shows that $\Tr(\mathbf{A})=\Tr(\mathbf{U})$, i.e. the anti-symmetric perturbation
$\mathbf{Q}\mathbf{U}$ does not contribute to the trace and therefore cannot change the exponential contraction rate
in \autoref{eq:mean_volume_contraction}.
Consequently, the collapse time $t_C$ defined via the entropic volume criterion is invariant under non-reversible perturbations $\mathbf{Q}$. The perturbation $\mathbf{Q}$ can reshape transient geometry and modify finite-time diagnostics, but it leaves the onset time unchanged because the leading contraction rate entering \autoref{eq:mean_volume_contraction} is fixed by $\Tr(\mathbf{A})=\Tr(\mathbf{U})$.

\subsection{Explicit Derivation for Diagonal Case}
We analyze the probability distribution $P_t^e(\mathbf{x})$ at a probe point $\mathbf{x} = e^{-\mathbf{U}t}\mathbf{a}_1 + \mathbf{z}_{\mathrm{sto}}$, where $\mathbf{z}_{\mathrm{sto}} \sim \mathcal{N}(0, \mathbf{\Sigma}_{\mathrm{sto}}(t))$. The distribution is a sum over all data points $\mu=1, \dots, n$:
\begin{align}
    P_t^e(\mathbf{x}) \propto \sum_{\mu=1}^{n} \exp\left(-\frac{1}{2} (\mathbf{x} - e^{-\mathbf{U} t} \mathbf{a}_\mu)^\top \mathbf{\Sigma}_{\mathrm{sto}}^{-1}(t) (\mathbf{x} - e^{-\mathbf{U} t} \mathbf{a}_\mu) \right).
\end{align}
The "energy" of the $\mu$-th term, $E'_\mu(t)$, is defined as the argument of the exponential multiplied by $-1$. Substituting the probe point $\mathbf{x}$, we get:
\begin{align}
    E'_\mu(t) = \frac{1}{2} \left(e^{-\mathbf{U}t}(\mathbf{a}_1 - \mathbf{a}_\mu) + \mathbf{z}_{\mathrm{sto}}\right)^\top \mathbf{\Sigma}_{\mathrm{sto}}^{-1}(t) \left(e^{-\mathbf{U}t}(\mathbf{a}_1 - \mathbf{a}_\mu) + \mathbf{z}_{\mathrm{sto}}\right).
\end{align}
Since all matrices are diagonal, this becomes a sum over components $j=1, \dots, d$:
\begin{align}
    E'_\mu(t) = \frac{1}{2} \sum_{j=1}^d \lambda_j(t) \left[e^{-u_j t}((\mathbf{a}_1)_j - (\mathbf{a}_\mu)_j) + (\mathbf{z}_{\mathrm{sto}})_j\right]^2,
\end{align}
where $\lambda_j(t) = u_j / (1-e^{-2u_j t})$. We assume the data differences $\delta_j = (\mathbf{a}_1)_j - (\mathbf{a}_\mu)_j$ are drawn from $\mathcal{N}(0, 2\sigma_0^2)$.

To calculate the free energy of the background terms ($\mu \neq 1$), we compute the expectation of $e^{-\beta E'_\mu(t)}$ over the distribution of $\mathbf{a}_\mu$. This decouples into a product of expectations over each component, $M_j(\beta) = \mathbb{E}_{\delta_j}[ \exp(-\frac{\beta}{2} \lambda_j(t) [e^{-u_j t}\delta_j + z_j]^2) ]$. This involves a standard Gaussian integral of the form $\int e^{-A\delta^2 - B\delta - C}d\delta$. The exponent is:
\begin{align}
    \mathcal{E} = -\delta_j^2 \left( \frac{1}{4\sigma_0^2} + \frac{\beta \lambda_j(t) e^{-2u_j t}}{2} \right) - \delta_j \left( \beta \lambda_j(t) e^{-u_j t} z_j \right) - \frac{\beta \lambda_j(t) z_j^2}{2}.
\end{align}
Solving this integral yields the result for $M_j(\beta)$:
\begin{align}
    M_j(\beta) = \left(1 + 2\sigma_0^2 \beta \lambda_j(t) e^{-2u_j t}\right)^{-1/2} \exp\left(-\frac{\beta \lambda_j(t) z_j^2}{2(1 + 2\sigma_0^2 \beta \lambda_j(t) e^{-2u_j t})}\right).
\end{align}
The free energy density is $g_t(\beta) = \frac{1}{d} \sum_j \ln M_j(\beta)$. In the large-$d$ limit, we replace the random variable $(\mathbf{z}_{\mathrm{sto}})_j^2$ with its typical (mean) value, $\mathbb{E}[(\mathbf{z}_{\mathrm{sto}})_j^2] = \sigma_j^2(t) = (\lambda_j(t))^{-1}$, which simplifies $\lambda_j(t)(\mathbf{z}_{\mathrm{sto}})_j^2 \approx 1$. This gives the final expression for the free energy density:
\begin{align}
    g_t(\beta) = \frac{1}{d} \sum_{j=1}^d \left[ -\frac{1}{2}\log\left(1 + 2\sigma_0^2 \beta \lambda_j(t) e^{-2u_j t}\right) - \frac{\beta}{2(1 + 2\sigma_0^2 \beta \lambda_j(t) e^{-2u_j t})} \right].
\end{align}
The collapse transition occurs when the log-density of the signal term, LHS $= -E'_1(t_C)/d \approx -1/2$, equals the log-density of the background terms, RHS $\approx \alpha + g_t(1)|_{t_C}$. Setting $\beta=1$ and LHS = RHS gives the condition $g_t(1)|_{t_C} = -\alpha - 1/2$. Substituting the expression for $g_t(1)$ and multiplying by $-2$ yields the final implicit equation for $t_C$, \autoref{eq:tc_rem}.

\end{document}